\documentclass{kluwer}
\newdisplay{guess}{Conjecture}
\begin{document}
\begin{article}
\begin{opening}
\tolerance=10000 \hbadness=10000
\def\verbfont{\small\tt} 
\def\note #1]{{\bf #1]}} 
\def\be{\begin{equation}}
\def\ee{\end{equation}} 
\def\bearr{\begin{eqnarray}} 
\def\eearr{\end{eqnarray}}
\def\barr{\begin{array}}

\title{P-MODES IN AND AWAY FROM A SUNSPOT}
\author{Brajesh \surname{Kumar}}
\author{Rajmal \surname{Jain}}
\author{S. C. \surname{Tripathy}}
\runningauthor{Brajesh Kumar et al.}
\runningtitle{P-MODES IN AND AWAY FROM A SUNSPOT}
\institute{Udaipur Solar Observatory, A unit of Physical Research Laboratory, Off Bari Road, Dewali, P.  B.  No.  198, Udaipur 313001, India} 
\author{Hari Om \surname{Vats}}
\author{M. R. \surname{Deshpande}}
\institute{Physical Research Laboratory, Navrangpura, Ahmedabad 380009, India} 
\date{22 October, 1999}

\begin{abstract}
A time series of GONG Dopplergrams for the period 10-14 May 1997 from Udaipur 
and Big Bear sites has been used to measure the velocity fluctuations in the 
sunspot (NOAA active region 8038) and quiet photosphere simultaneously. We observe 
that the power of pre-dominant p-mode is reduced in the sunspot as compared 
to quiet photosphere by 39-52$\%$ depending on the location of the sunspot 
region on the solar disk. We also observe a relative peak frequency deviation 
of p-modes in the sunspot, of the order of 80-310 $\mu$Hz, which shows a 
linear dependence on the magnetic field gradient in the active region. The 
maximum frequency deviation of 310 $\mu$Hz on 12 May appears to be an influence of a long
duration solar flare that occurred in this active region. We interpret this 
relative peak frequency deviation as either due to power re-distribution 
of p-modes in the sunspot or a consequence of frequency modulation of these 
modes along the magnetic flux tubes due to rapidly varying magnetic field 
structure.
\end{abstract}

\end{opening}

\section{Introduction} 
The study of intensity and velocity variations in and above an active
region is continuously refined with successively improved observational
techniques. The intent of such studies is to understand the basic
mechanism of the interaction of acoustic waves with the complex magnetic
features. These investigations are carried out by measuring and
quantifying the scattering properties viz. amplitude, frequency and phase
of solar eigen-modes. There are two possible methods to carry out this
study. The first method involves the relative study of the properties of
solar surface waves travelling towards an active region and the wave that
is reflected from it (Tarbell {\it et al.} 1988; Braun, Duvall, Jr., and LaBonte, 
1988; Braun {\it et al.} 1992; Bogdan {\it et al.} 1993; Keppens, 1994; Cally, 1995). 
The other way can be a direct observation in and
above an active region at photospheric and chromospheric layers
respectively, which relates to the possible modulation of these acoustic
waves along the magnetic flux tubes (Giovanelli and Slaughter, 1978; 
Lites, White, and Packman, 1982; Title {\it et al.} 1992; Hindman and Brown, 1998; 
Balthasar {\it et al.} 1998). The investigations by both the methods
have revealed a reduction in p-mode amplitude within active regions due to
absorption of acoustic waves by magnetic structures. It is well established 
that the typical sunspots absorb as much as 40-50$\%$ of the power of incident 
acoustic waves (Braun, Duvall, Jr., and LaBonte, 1987). However, Braun and 
Duvall Jr. (1990) observed power reduction of nearly 70$\%$ in a large 
sunspot group (NOAA active region 5395).

There has been ample evidence towards significant deviations in peak frequency 
of the Doppler velocity and intensity oscillations in sunspots at upper 
photospheric and chromospheric layers with the help of a lot of new 
measurements in different spectral lines and with improved observational 
techniques (Kentischer and Mattig, 1995 and references therein). On the 
contrary, the p-mode oscillations in sunspots at photospheric layers have 
been reported to be of 300s (Beckers and Schultz, 1972; Rice and Gaizauskas, 
1973; Soltau, Schroter, and Wohl, 1976; Livingston and Mahaffey, 1981) and 
310 sec (Bhatnagar, Livingston, and Harvey, 1972), which is 
nearly same as that in the quiet photosphere. 
However, Nye {\it et al.} (1981) showed that the mean power spectrum of the 
umbral oscillations has several clear peaks, including a group of three 
peaks clustered around five minute period (260s, 300s and 350s) and a peak 
at about 200s. Similarly, Horn, Staude, and Landgraf (1997) have shown that the spectra 
of velocity oscillations in the sunspot show features of strengthened power 
peaks in the bands of periods around 3 min and weakened in 5 min with 
respect to the quiet Sun. 

In this paper, a time series of GONG Dopplergrams has been analysed to 
improve our present understanding of the influence 
of magnetic fields on the oscillation properties of acoustic waves near a 
sunspot. The GONG instrument provides a spatial resolution of 8
arc-sec/pixel, which may be taken as a reasonable resolution for 
comparative study of p-mode oscillations in sunspot and quiet regions.

\section{Data Analysis and Results}
The data used in this study consists of Udaipur and Big Bear GONG
Dopplergrams taken during 10-14 May 1997. The GONG instrument 
(Harvey {\it et al.} 1996) uses Ni I 6768 Angstrom spectral line to measure 
the global intensity and velocity variations in the solar photosphere. 
This instrument acquires
full disk images of the Sun on a 256x256 pixel CCD camera implying
a spatial resolution of 8 arc-sec/pixel. In spite of this low
spatial resolution and a constant parasitic stray light problem in the
sunspot, the GONG data can be reasonably used to investigate the influence
of magnetic field on the propagation of p-modes by carrying out a
comparative study in quiet and sunspot region. 
We have used once a minute GONG Dopplergrams of the observing interval on
each day during 10-14 May to measure the velocity fluctuations in the
sunspot (NOAA active region 8038) and quiet photosphere simultaneously
on the solar disk. The observing interval along with the position of the
sunspot in Heliographic coordinates is given in Table~1. 
We have carried out this analysis for five continuous days, and also 
included data from two different stations, to enhance the statistical confidence.

\begin{table*} \caption[]{Data Description}
\label{tab1}
\begin{tabular}[]{lccc}
\hline 
\noalign{\smallskip} 
Station name & Date & Observing Time & Average sunspot location \\
&  (1997) & (UT) & (Heliographic Coordinates) \\
\noalign{\smallskip} 
\hline
Udaipur & 10 May & 03:30-11:00 & N20E17 \\
&  11 May & 02:16-12:36 & N21E04 \\
&  12 May & 02:30-11:30 & N21W10 \\
&  13 May & 03:00-11:30 & N21W22 \\
&  14 May & 03:00-11:20 & N21W35 \\
Big Bear & 12 May & 15:00-20:20 & N21W16 \\
 & 14 May & 15:00-23:20 & N21W41 \\
\noalign{\smallskip}
\hline
\end{tabular} 
\end{table*}

The intensity and velocity images of the solar disk for 12 May taken at 05:13 
UT from Udaipur site are shown in Figures~1 and 2 respectively. Since a sunspot is not 
visible on the Dopplergram, the corresponding intensity filtergrams 
have been used to locate the position of the sunspot on the solar disk.
It is well known that Dopplergrams in addition to p-modes also
exhibit variety of features such as supergranulation pattern, meridional 
flows and solar rotation gradients. A two point backward difference 
filter (GRASP/IRAF software package), 
\begin{equation}
	Filtered Image(t) = Image(t) - Image(t-1)
\end{equation}
is applied to the Dopplergrams to enhance the p-mode oscillations above 
other features.

The size of the sunspot under study is about 20 arc-sec, which is 
three times higher than the spatial resolution of GONG instrument. The
size of the sunspot remains nearly the same for the period under study, and
hence it remains confined within a pixel matrix of 3x3 during its passage
on the solar disk during 10-14 May. Thus we chose a grid of 3x3 pixels for
measuring the velocity fluctuations inside the sunspot in the time series
of each individual day. On the other hand, three quiet photospheric
regions (q1, q2 and q3) of equal grid size have been chosen at
approximately the same limb distance to account for the
effect of limb darkening, foreshortening and various other known
effects. The selection of three quiet regions, instead of a single region
on the opposite longitude, is carried out in view of 
(i) to increase the statistical confidence for a
comparative study and (ii) to quantify the variation in the
properties of acoustic modes in different locations of the quiet
photosphere.

The velocity fluctuations are measured for each pixel inside the grid as a
function of time. In Figure~3, we illustrate these fluctuations for 12 May 
1997, Udaipur station. This time series of each pixel 
is subjected to Fourier Transform to obtain the power spectrum. The power
spectra from 9 pixels in the grid are then averaged to improve the signal
to noise ratio as suggested earlier by Kentischer and Mattig (1995). In
this way, we obtain four average power spectra for each day as
shown by dashed lines in Figures~4-10 . It may be noticed that the power 
in sunspot is appreciably reduced as compared to
all the quiet regions while the power and frequency of
p-modes in quiet regions do not vary with statistical significance.

We notice that the power spectrum of acoustic modes in the quiet region 
follows an asymptotic
Lorentzian distribution as observed earlier by Anderson {\it et al.} (1990).
On the other hand, a significant departure from such a Lorentzian profile
can be seen in the sunspot power spectrum. The power spectrum of p-mode
oscillations in sunspot shows distinctly several peaks around 5 min period
in agreement to earlier investigations (Nye {\it et al.} 1981; Horn, Staude, and
Landgraf, 1997) which have been obtained using high resolution
observations. Based on the approach for modelling the solar oscillation
spectra for global p-modes, Anderson {\it et al.} (1990) suggested that Lorentzian
fitting could be a suitable peak finding algorithm. However, in view of our
sunspot results, the Lorentzian fitting may not be the best peak finding 
technique. Also with the aim of 
comparing the properties of p-mode oscillations between 
sunspot and quiet region, we need to apply the same peak finding algorithm 
to both the regions. Thus, to determine the
genuine peak power and corresponding frequency of the power envelope, 
we applied a low pass digital filter, namely a Savitzky-Golay filter (Press
{\it et al.} 1992), to the average power spectrum for both quiet and sunspot
regions separately. This filter basically smooths the data by a window
function of a predefined number of data points and a polynomial order with a
proper weighting. After several experimental values of 
data points and polynomial orders, the optimal statistical fit is 
observed with a window of 32 data points and a polynomial of order 6. We 
have then applied it to the average power spectrum for each day during 10-14 
May to obtain the proper fitted power envelope as shown in Figures~4 to 10 by 
solid lines.

\begin{table*} \caption[]{Power and frequency estimate from S-G filter for
quiet regions (q1, q2 $\&$ q3) and sunspot}
\label{tab2}
\begin{tabular}[]{lcccccccc}
\hline 
\noalign{\smallskip} 
Date & $P_{q1}$ & $\nu_{q1}$ & $P_{q2}$ & $\nu_{q2}$ & $P_{q3}$ & $\nu_{q3}$ & $P_{s}$ & $\nu_{s}$ \\
(1997) & (1E5) & (mHz) & (1E5) & (mHz) & (1E5) & (mHz) & (1E4) & (mHz) \\
\noalign{\smallskip} 
\hline
Udaipur data: & & & & & & & & \\
10 May & 1.63 & 3.26 & 1.65 & 3.26 & 1.67 & 3.30 & 8.79 & 3.19 \\
11 May & 1.86 & 3.26 & 1.90 & 3.29 & 1.89 & 3.26 & 11.27 & 3.18 \\
12 May & 1.74 & 3.26 & 1.77 & 3.29 & 1.76 & 3.26 & 9.95 & 2.96 \\
13 May & 1.63 & 3.25 & 1.68 & 3.28 & 1.65 & 3.28 & 9.28 & 3.09 \\
14 May & 1.57 & 3.25 & 1.58 & 3.28 & 1.59 & 3.25 & 7.58 & 3.14 \\
Big Bear data: & & & & & & & & \\
12 May & 1.03 & 3.26 & 1.04 & 3.32 & 1.05 & 3.26 & 6.39 & 3.07 \\
14 May & 1.24 & 3.26 & 1.27 & 3.26 & 1.27 & 3.29 & 6.28 & 3.15 \\
\noalign{\smallskip}
\noalign{\smallskip}
\hline
\end{tabular} 
\end{table*}

\begin{table*} \caption[]{Estimation of relative power reduction and frequency
deviation in the sunspot}
\label{tab3}
\begin{tabular}[]{lcccccccc}
\hline 
\noalign{\smallskip} 
Date & $P_{q}$ & $\nu_{q}$ & $P_{s}$ & $\nu_{s}$ & $\delta\nu$ & $\Delta$P & $\Delta\nu$ & $\Delta\nu$ \\
(1997) & (1E5) & (mHz) & (1E4) & (mHz) & ($\mu$Hz) & ($\%$) & ($\mu$Hz) & ($\sigma$) \\
\noalign{\smallskip} 
\hline
Udaipur data: & & & & & & & & \\
10 May & 1.65 & 3.27 & 8.79 & 3.19 & 34.72 & 46.66 & 80 & $>$2$\sigma$ \\
11 May & 1.88 & 3.27 & 11.27 & 3.18 & 26.88 & 40.15 & 90 & $>$3$\sigma$ \\
12 May & 1.76 & 3.27 & 9.95 & 2.96 & 30.80 & 43.39 & 310 & $>$10$\sigma$ \\
13 May & 1.66 & 3.27 & 9.28 & 3.09 & 32.68 & 43.97 & 180 & $>$5$\sigma$ \\
14 May & 1.58 & 3.26 & 7.58 & 3.14 & 33.33 & 52.01 & 120 & $>$3$\sigma$ \\
Big Bear data: & & & & & & & & \\
12 May & 1.04 & 3.28 & 6.39 & 3.07 & 52.00 & 38.52 & 210 & $>$4$\sigma$ \\
14 May & 1.26 & 3.27 & 6.28 & 3.15 & 33.33 & 50.13 & 120 & $>$3$\sigma$ \\
\noalign{\smallskip}
\noalign{\smallskip}
\hline
\end{tabular} 
\end{table*}

The estimated peak power and the corresponding frequency in 
the quiet regions (q1, q2 $\&$ q3) and sunspot are shown in Table~2 for 
each individual day. Based on the results listed in Table~2, we have 
calculated the variation in peak power and the corresponding frequency 
of the power envelopes for the three quiet regions. It is observed that 
the variation in power among q1, q2 and q3 is within 5$\%$,
whereas the variation in frequency remains within the frequency resolution
limit ($\delta\nu$) of the power spectra. This implies that the variation in power and
frequency of the power envelope in the different quiet regions is within
the limits of one standard deviation. Thus we conclude that the
properties of p-modes do not vary significantly from one quiet region
to other within the error limits. Therefore, we have averaged the
peak power and corresponding frequency of all the quiet regions, which
is then compared with that of the sunspot. 
The peak power ($P_{s}$) and the corresponding frequency 
($\nu_{s}$) in the sunspot with the average peak power ($P_{q}$) and 
frequency ($\nu_{q}$) in the quiet regions are given in Table~3. It is 
observed that the sunspot peak power is reduced ($\Delta$P) by 39-52$\%$ as compared to
quiet region during its passage on the solar disk. It is
interesting to note that the corresponding frequency of the power
envelope in the sunspot deviates ($\Delta\nu$) in the range of 80-310 $\mu$Hz
relative to the quiet region. The comparison
of the power envelopes estimated by S-G filter for quiet and sunspot
regions for Udaipur station on 12 May is shown in Figure~11. 
The relative power reduction in the sunspot is clearly seen. A relative 
frequency shift of the power envelope in the sunspot is also noted.

\section{Discussion and Conclusions}
Our results distinctly show the power reduction in the sunspot relative
to a quiet region on each day in agreement to earlier
investigations. However, the amount of relative reduction in
power in the sunspot ($\Delta$P) is found to be dependent on the location 
of the sunspot. During the period 10-13 May, when the sunspot is
found within $\pm$$25\deg$ to the central meridian (Table~1), $\Delta$P varies
around 43$\%$. On the other hand, on 14 May, when the sunspot moves further 
apart from the central meridian, $\Delta$P is found to be significantly higher;
of the order of 52$\%$ as illustrated in Table 3.
Our study also shows an apparent frequency deviation of the power
envelope of acoustic modes in the  sunspot as compared to quiet regions.
This frequency deviation ($\Delta\nu$), as shown in Table 3, is also varying during 
10-14 May, the lowest being on 10 May and the highest on 12 May.  
It may be noted from Table~3 that the observed frequency
deviation varies from 2-10$\sigma$, where 1$\sigma$
is taken equivalent to the frequency resolution ($\delta\nu$) of the power spectrum. 
The frequency deviation of the power envelope ($\Delta\nu$) is considered to 
be of statistical significance when $\Delta\nu$ $\geq$ 3$\sigma$. Following 
this, we find that only the deviation between
11-14 May is significant. As observed from the Udaipur data,
$\Delta\nu$ increases from 2$\sigma$ on 10 May to 10$\sigma$ on 12 May
and then drops to 5$\sigma$ and 3$\sigma$ on 13 May and 14 May respectively.
Recently, Jain {\it et al.} (1999), based on the study of evolution of a sunspot
region using SOHO magnetograms, have shown that the magnetic structure 
in this active region was rapidly varying. This conclusion was inferred 
from the growth and decay of emerging flux of both
polarities close to the sunspot. As a ready reference, we show 
a sequence of few high resolution magnetograms 
of this active region taken from Jain {\it et al.} (1999) in Figure~12.  
It is observed that two major opposite polarities were approaching towards 
each other since 10 May, resulting in the growth of the magnetic field gradient, 
so as to collide around 04:42 UT on 12 May leading to a long duration solar 
flare event (04:42-06:20 UT) of 1B importance and an associated
CME. We have calculated the daily variation in magnetic
field gradient of this active region between 10-14 May and plotted it with
the daily variation in peak frequency deviation in the sunspot (see Figure 13). 
This plot shows a strong linear dependence of the frequency
deviation on the magnetic field gradient and hence magnetic structure
of the active region. This indicates
the role of the magnetic field gradient on the frequency deviation of the
peak power in the power envelope of the acoustic modes of the sunspot
relative a quiet region. The extraordinary increase in $\Delta\nu$,
of the order of 310 $\mu$Hz, on 12 May for Udaipur station may
be interpreted as the influence of the long duration solar flare in the course of
observing sequence as detected earlier by Jain and Tripathy (1998) while
investigating the chromospheric modes in solar flare using high resolution
H$\alpha$ filtergarms. However, we obtain a frequency shift of 
210 $\mu$Hz on 12 May from the Big Bear data comprising of a post flare 
time slot (15:00-20:00 UT), which can be understood by the simplification of 
the magnetic field structure as explained by Jain {\it et al.} (1999).  
The frequency deviation obtained from Big Bear data on 14 May is consistent 
with that obtained from Udaipur site on this day.

We conclude that in addition to the magnetic field gradient, the high energy 
build-up processes in the solar
flares influence the properties of p-modes in the sunspot as compared to
the quiet region. We conjecture that the deviation of the peak frequency of
power envelope of p-modes in the sunspot as compared to quiet region is
either due to a power re-distribution so as to peak at nearby
frequency, or it could be the frequency modulation of these acoustic modes 
along the magnetic flux tube underneath the sunspot.

\acknowledgements
We are thankful to A. Bhatnagar, R. Muller, H. M. Antia and P.
Venkatakrishnan for useful discussions and suggestions. We also wish
to express our thanks to Kiran Jain for her help in the data analysis
work. Our sincere thanks to W. C. Livingston for his critical 
comments and suggestions which improved the manuscript.
This work utilizes data obtained by the Global Oscillation Network
Group (GONG) project, managed by the National Solar Observatory, a
Division of the National Optical Astronomy Observatories, which is
operated by AURA, Inc. under a cooperative agreement with the National
Science Foundation. The data were acquired by instruments operated by
the Big Bear Solar Observatory and Udaipur Solar Observatory.

\newpage
\begin{figure}
\begin{center}
\leavevmode
\caption{Calibrated GONG intensity filtergram (IZI) of the solar disk
acquired at Udaipur site on 12 May 1997 at 05:13 UT. The image has been
corrected for camera offset, camera rotation and camera pixel aspect
ratio. The location of the sunspot is marked.} 
\end{center}
\end{figure}

\begin{figure}
\begin{center}
\leavevmode
\caption{Calibrated GONG Dopplergram (VZI) of the solar disk acquired
at Udaipur site on 12 May 1997 at 05:13 UT. The image has been corrected for
camera offset, camera rotation and camera pixel aspect ratio and then passed
through a two point backward difference filter to enhance the p-mode
oscillations above other features.}
\end{center}
\end{figure}

\begin{figure}
\begin{center}
\leavevmode
\caption{Time series of velocity oscillations of a single pixel for quiet
and sunspot region on 12 May 1997 (02:30-11:30 UT), Udaipur station.} 
\end{center}
\end{figure}

\begin{figure}
\begin{center}
\leavevmode
\caption{Average power spectra (dashed lines) of p-modes over a grid of
3x3 pixels for quiet regions (q1, q2 $\&$ q3) and sunspot on 10 May 1997 for
Udaipur station. Shown in solid line is fit estimated by S-G filter after smoothing
the data by a window function of 32 points and a polynomial of order 6.} 
\end{center}
\end{figure}

\begin{figure}
\begin{center}
\leavevmode
\caption{Same as Figure 4 for 11 May 1997, Udaipur station.} 
\end{center}
\end{figure}

\begin{figure}
\begin{center}
\leavevmode
\caption{Same as Figure 4 for 12 May 1997, Udaipur station.} 
\end{center}
\end{figure}

\begin{figure}
\begin{center}
\leavevmode
\caption{Same as Figure 4 for 13 May 1997, Udaipur station.} 
\end{center}
\end{figure}

\begin{figure}
\begin{center}
\leavevmode
\caption{Same as Figure 4 for 14 May 1997, Udaipur station.} 
\end{center}
\end{figure}

\begin{figure}
\begin{center}
\leavevmode
\caption{Same as Figure 4 for 12 May 1997, Big Bear station.} 
\end{center}
\end{figure}

\begin{figure}
\begin{center}
\leavevmode
\caption{Same as Figure 4 for 14 May 1997, Big Bear station.} 
\end{center}
\end{figure}

\begin{figure}
\begin{center}
\leavevmode
\caption{Comparison of the power envelopes estimated by S-G filter for quiet
(solid line) and sunspot (dashed line) for Udaipur station on 12 May 1997. The
relative power reduction in the sunspot is clearly seen. A relative frequency
shift of the power envelope in the sunspot may also be noted.} 
\end{center}
\end{figure}

\begin{figure}
\begin{center}
\leavevmode
\caption{Sequence of a few high resolution magnetograms of NOAA active region 8038
obtained by MDI/SOHO for the period 11-13 May 1997 (Jain et al. 1999). The relative motion of the
regions of opposite polarities can be clearly understood from this sequence of
magnetograms of the active region. The ejection of the north polarity flux
from the sunspot and growth of new EFRs are observed. The collision of
opposite polarities took place at around 04:42 UT on 12 May 1997 and is also 
visible in the magnetogram at 04:52 UT. This led to a
long duration solar flare event (04:42-06:20 UT) of 1B importance.}
\end{center}
\end{figure}

\begin{figure}
\begin{center}
\leavevmode
\caption{Plot showing the variation of peak frequency deviation of the power
envelope of p-modes in the sunspot with the change in magnetic field gradient
of the sunspot active region. The error bars indicate 1$\sigma$ value.}
\end{center}
\end{figure}

\end{article}
\end{document}